\documentclass[twocolumn,showpacs,preprintnumbers,amsmath,amssymb]{revtex4}

\def\ube13{UBe$\rm_{13}$}

\def\bi2212{Bi$\rm_2$Sr$\rm_2$CaCu$\rm_2$O$\rm_8$}
\def\ybi2212{Bi$\rm_2$Sr$\rm_2$YCu$\rm_2$O$\rm_8$}
\def\ycabi2212{Bi$\rm_2$Sr$\rm_2$Ca$\rm_{1-x}$Y$\rm_x$Cu$\rm_2$O$\rm_{8+\delta}$}

\def\y65cabi2212{Bi$\rm_2$Sr$\rm_2$Ca$\rm_{0.35}$Y$\rm_{0.65}$Cu$\rm_2$O$\rm_{8+\delta}$}

\def\Co{CeCoIn$_5$}

\usepackage{graphicx}
\usepackage{dcolumn}
\usepackage{bm}

\begin{document}

\preprint{submitted to Physical Review Letters}

\title{Fulde-Ferrell-Larkin-Ovchinnikov Superconducting State in CeCoIn$_5$}

\author{A. Bianchi, R. Movshovich, C. Capan, A. Lacerda, P. G. Pagliuso, and J. L. Sarrao}
\affiliation{%
Los Alamos National Laboratory, Los Alamos, New Mexico 87545
}%


\date{\today}

\begin{abstract}

We report specific heat measurements of the heavy fermion superconductor \Co\ in the vicinity of
the superconducting critical field $H_{c2}$, with magnetic field in the [110], [100], and [001]
directions, and at temperatures down to 50 mK. The superconducting phase transition changes from
second to first order for field above 10 T for $H \parallel$ [110] and $H \parallel$ [100]. In the
same range of magnetic field we observe a second specific heat anomaly within the superconducting
state. We interpret this anomaly as a signature of a Fulde-Ferrell-Larkin-Ovchinnikov (FFLO)
inhomogeneous superconducting state. We obtain similar results for $H \parallel$ [001], with FFLO
state occupying a smaller part of the phase diagram.

\end{abstract}

\maketitle

In the early 1960's, following the success of the BCS theory of superconductivity, Fulde and
Ferrell~\cite{fulde-ferrell:pr-64} and Larkin and Ovchinnikov~\cite{larkin-ovchinnikov:jetp-64}
developed theories of inhomogeneous superconducting states. At the core of FFLO theory lie
competing interactions of a very basic nature. One is the interaction of the spin of the electron
with magnetic field and the other is the energy of the superconducting coupling of electrons into
Cooper pairs, or the condensation energy. In the normal state the electrons are free to lower their
total energy by preferentially aligning their spins along the magnetic field, leading to a
temperature-independent Pauli susceptibility. For spin-singlet superconductors (both {\it s-} and
{\it d-wave}), the condensate contains an equal number of spin-up and spin-down electrons.
Therefore, Pauli paramagnetism will always favor the normal state over the spin-singlet
superconducting state, and will reduce the superconducting critical field $H_{c2}$ which suppresses
superconductivity. This effect is called Pauli limiting, with the characteristic Pauli field $H_P$
determining the upper bound of $H_{c2}$~\cite{clogston:prl-62}. Another effect of magnetic field
that leads to the suppression of superconductivity is orbital limiting, or suppression of
superconductivity when the kinetic energy of the supercurrent around the normal cores of the
superconducting vortices in Type II superconductors becomes greater than the superconducting
condensation energy. The orbital limiting field $H_{c2}^0$ defines $H_{c2}$ in the absence of Pauli
limiting. The relative strength of Pauli and orbital limiting, the so called Maki parameter $\alpha
= H_{c2}^0/H_P$, determines the behavior of the system in high magnetic field. The prediction of
FFLO theory is that for a clean Type II superconductor with sufficiently large $\alpha$ (for
$\alpha > 1.8$ in the calculations of Ref.~\onlinecite{gruenberg:prl-66}), a new inhomogeneous
superconducting FFLO state will appear between the normal and the mixed, or vortex, state below the
critical temperature $T_{FFLO}$~\cite{gruenberg:prl-66}. Within the particular realization of
Larkin and Ovchinnikov~\cite{larkin-ovchinnikov:jetp-64}, this state is characterized by the
appearance of a periodic array of planes of normal electrons that can take advantage of the Pauli
susceptibility.

A number of conventional superconductors were proposed as candidates for observation of the FFLO
state, due to their high orbital critical field $H_{c20}$ and, therefore, relatively strong Pauli
limiting effect, in the early and mid-sixties. Experimental searches, however, yielded null
results~\cite{berlincourt:pr-63,kim:pr-65,shapira:pr-65,hake:prl-65}. The failure to observe the
FFLO state was attributed to high spin-orbit scattering rate in these compounds~\cite{maki:pr-66}.
In the last decade the FFLO state was suggested to exist in heavy fermion UPd$_2$Al$_3$
(Ref.~\cite{gloos:prl-93} and CeRu$_2$ (Ref.~\cite{huxley:jpcm-93}), based on thermal expansion and
magnetization data, respectively. Subsequent research identified the magnetization feature in
CeRu$_2$ as due to flux motion~\cite{tenya:physicab-99}, and the region of the suggested FFLO state
in UPd$_2$Al$_3$ was shown to be inconsistent with theoretical models~\cite{norman:prl-93}. Most
notably, multiple phase transitions that can be associated with the FFLO state have not been
observed with a single measurement technique.

Heavy-fermion superconductor \Co\ satisfies all requirements of theory for the formation of the
FFLO state. It is very clean, with an electronic mean free path on the order of microns in the
superconducting state, which significantly exceeds the superconducting correlation
length\cite{movshovich:prl-01}. Its Maki parameter $\alpha \approx 3.5$ is twice the minimum
required for the formation of the FFLO state~\cite{bianchi:prl-02}. It was recently discovered that
the superconducting phase transition changes from second to first order at $T_0 \approx 0.3 T_c$
for field $H
\parallel $ [001], which was taken as an indication that Pauli limiting drives the physics of \Co\
at low temperature and high magnetic field~\cite{bianchi:prl-02,maki:ptp-64}. The critical point
$T_0$ was found to be in very good agreement with the one predicted by FFLO theory for a compound
with $\alpha = 3.5$~\cite{gruenberg:prl-66}. Magnetization measurements of Tayama {\it et
al.}~\cite{tayama:prb-02} showed that the superconducting transition in \Co\ becomes first order at
a critical temperature $T_{0} = 0.7$ K for both $H \parallel$ [001] and $H
\parallel$ [100]. Magnetization measurements of Murphy {\it et al.}~\cite{murphy:prb-02} with $H
\parallel$ [110] indicated the presence of a second temperature-independent, $H \approx $ 8 T, anomaly
below 1.4 K, and the authors suggested that these results were consistent with FFLO state.

Materials with quasi-two-dimensional Fermi surfaces, which are likely to exhibit Fermi surface
nesting, are expected to have more stable FFLO phases when magnetic field lies within the 2D-like
planes~\cite{shimahara:prb-94}. De Haas-van Alphen studies of \Co\ revealed that a part of its
Fermi surface is an undulating cylinder with the axis along the (001) direction, characteristic of
the quasi-two-dimensional systems with planes perpendicular to [001]~\cite{hall:prb-01}.  These
theoretical~\cite{shimahara:prb-94} and experimental observations motivated us to perform specific
heat investigation of \Co\ with magnetic field $H \perp$ [001].

Specific heat data were collected by employing two techniques: the standard quasi-adiabatic method
and the temperature decay method, where a complete specific heat data set for a given field was
obtained by differentiating a single temperature versus time curve, generated as the sample was
coming into equilibrium with the bath starting from high temperature (above 1 K). This technique
was employed previously to resolve a sharp specific heat anomaly associated with the first order
superconducting phase transition in \Co\ for $H \parallel$ [001]~\cite{bianchi:prl-02}, and was
demonstrated to give high resolution data consistent with the quasi-adiabatic method.

\begin{figure}
\includegraphics[width=3in]{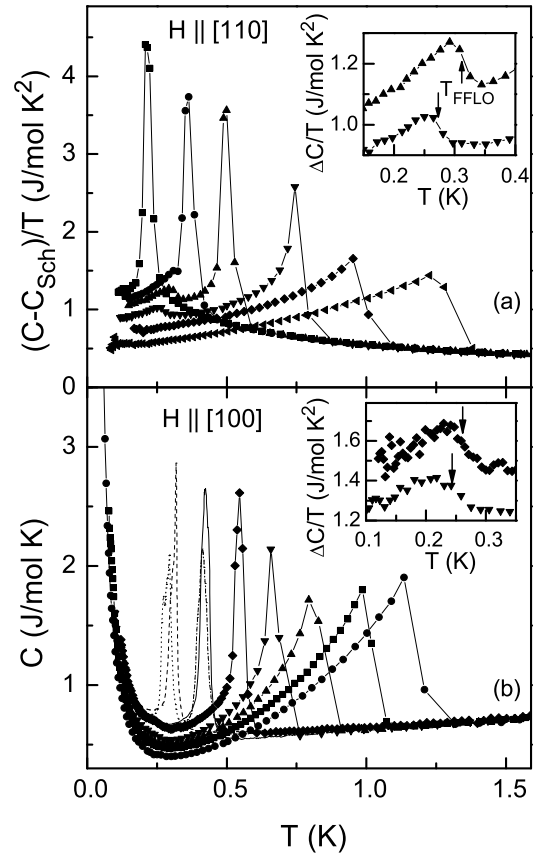}
\caption{ Specific heat of \Co\ with $H \perp $ [001]. (a) $H \parallel$ [110] data for fields of 9
T, 10 T, 10.6 T, 11 T, 11.2 T, and 11.4 T from right to left, collected with heat pulse method.
Inset: Low temperature region for 10.6 T and 11 T (same symbols) emphasizing the $T_{FFLO}$
anomaly. Arrows indicate phase transition temperatures from equal area construction. (b) $H
\parallel $ [100] Solid symbols: heat pulse data for fields of 9.5 T, 10 T, 10.5 T, 10.8 T, and 11
T from right to left. Solid (dash-dotted) curve is for 11.2 T data collected with the decay method
with temperature swept up (down). Dashed (dotted) curve is for 11.4 T with temperature swept up
(down). Inset: $T_{FFLO}$ anomalies for 10.8 T and 11 T.} \label{hc-pulsed}
\end{figure}

Figure~\ref{hc-pulsed} shows specific heat data of \Co\ collected with the quasi-adiabatic method,
as Sommerfeld coefficient $\gamma = C/T$ after subtraction of the Schottky anomaly tail at low
temperature, due to In and Co nuclear levels~\cite{movshovich:prl-01}, for magnetic field $H
\parallel$ [110] (panel (a)), and specific heat for $H \parallel $ [100] (panel (b)), as a function
of temperature. The superconducting anomaly at lower fields $H \le 10$ T is mean-field-like, with a
step in the specific heat at $T_c$, similar to the case of $H \parallel$ [001] when the field is
far from $H_{c2}$~\cite{bianchi:prl-02}. In this range increasing magnetic field simply reduces the
magnitude of anomaly, without changing the character of the transition. As the field is increased
further, the trend changes dramatically: the magnitude of the anomaly in the specific heat starts
to increase, and the anomaly itself sharpens up and acquires symmetric character, characteristic of
first order phase transitions. The specific heat data indicate that the change from second to first
order occurs at a critical magnetic field $H_0 \approx 10$ T and a critical temperature $T_0
\approx 1$ K. As the superconducting transition temperature is suppressed by the magnetic field
below $\approx 500$ mK, the transition becomes hysteretic (the data for 11.2 T and 11.4 T in
Fig.~\ref{hc-pulsed}(b)), proving unambiguously that the superconducting transition in \Co\ at high
fields close to the critical field $H_{c2}$ is indeed first order. At a temperature of about 300 mK
the specific heat data displays an additional anomaly within the superconducting state for $H \ge
10$ T, which we call a $T_{FFLO}$ anomaly. The low temperature region, in the vicinity of the
$T_{FFLO}$ anomaly, is shown in the insets of Figures~\ref{hc-pulsed}(a) and (b), where
$T_{FFLO}$'s for different fields are indicated by the arrows. The $T_{FFLO}$ anomaly can be
described as a step followed by a gradual decrease of the specific heat as with decreasing
temperature, a behavior characteristic of the second order phase transition. The $T_{FFLO}$ anomaly
is observed only in the superconducting state, and disappears when the superconducting phase
transition is suppressed by magnetic field below $T_{FFLO}$, as illustrated by the data for $H =
11.4$ T in Fig.~\ref{hc-pulsed}(a), or when $H \le 10$ T.

\begin{figure}
\includegraphics[width=3.5in]{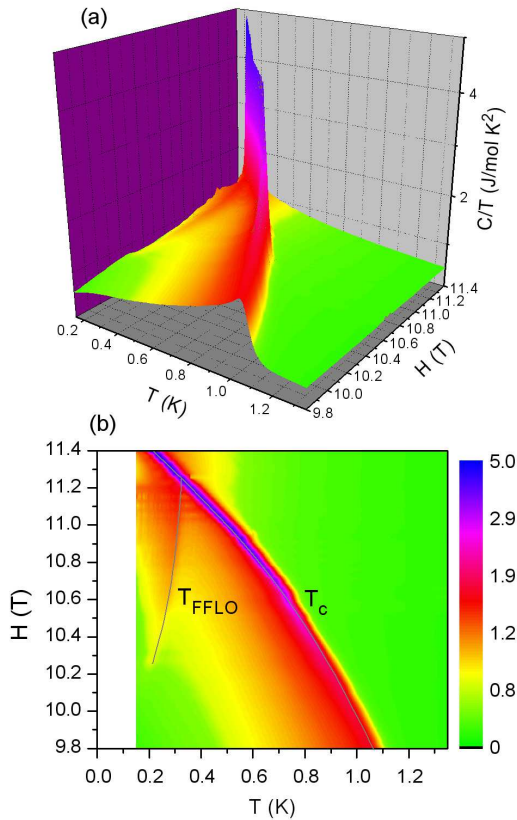}
\caption{(a) Electronic specific heat of \Co\ divided by temperature with $H \parallel $ [110]
collected with the temperature decay method, as a function of field and temperature. (b) Contour
plot of the data in (a) in the $H-T$ plane. Grey lines indicate the superconducting phase
transition $T_c$ and the FFLO-mixed state $T_{FFLO}$ anomaly. Color scale is the same in (a) and
(b).} \label{hc-color}
\end{figure}

The specific heat data collected with the decay method for $H \parallel$ [110] is displayed in
Fig.~\ref{hc-color}(a) as a surface contour plot in the $H-T$ plane. We can see a clear evolution
of the character of the specific heat anomaly with increasing magnetic field from a mean-field-like
step to a very sharp peak at higher magnetic field, as well as the development of the second low
temperature $T_{FFLO}$ anomaly (a red ridge) in the low temperature/high field corner of the H-T
plane. By plotting the data as color contour plot in Fig.~\ref{hc-color}(b) we can immediately
obtain the low temprerature/high field part of the phase diagram of \Co\ with $H \parallel$ [110],
where both superconducting - normal phase boundary $T_c$ and the $T_{FFLO}$ anomalies are indicated
by gray curves.

\begin{figure}
\includegraphics[width=3in]{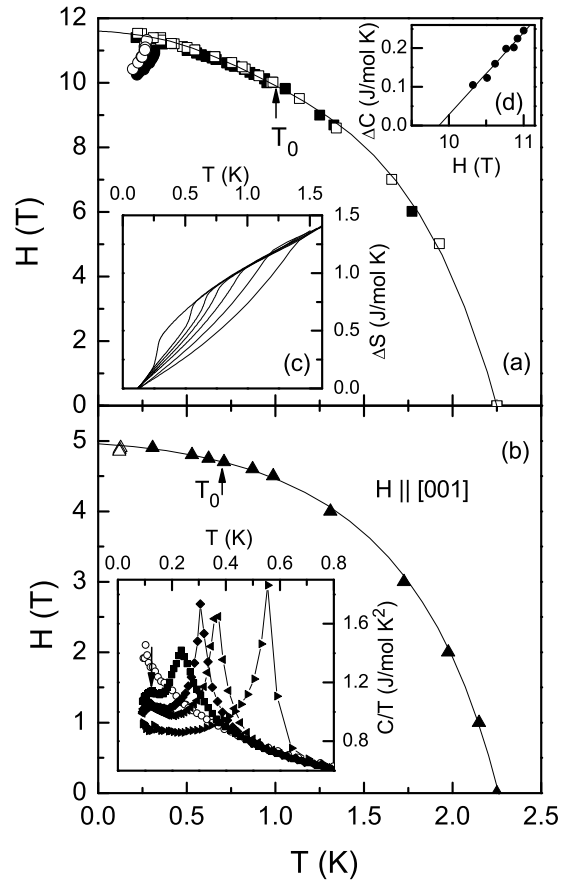}
\caption{(a): $H-T$ phase diagram of \Co\ with both $H \parallel $ [110] (filled symbols) $H
\parallel$ [100] (open symbols). ($\circ$) and ($\bullet$) indicate the $T_FFLO$ anomaly for $H
\parallel$ [100] and $H \parallel$ [110], respectively. Inset (c): Entropy gain from $T = 0.13$
K for fields of 11.4 T, 11 T, 10.8 T, 10.6 T, 10.22 T, 9.5 T, and 8.6 T from left to right. Inset
(d): Specific heat jump at the $T_{FFLO}$ anomaly obtained from equal area construction. (b): $H-T$
phase diagram for $H \parallel$ [001]. ($\triangle$) indicate $T_{FFLO}$ anomaly. Inset: Sommerfeld
coefficient; ($\circ$) 5T, solid symbols: 4.9 T, 4.875 T, 4.85 T, and 4.8 T from left to right.
Arrow indicates $T_{FFLO}$ anomaly at 4.9 T.  Solid lines in (a) and (b) are guides to eye for
superconducting phase boundaries.} \label{phase-diagram}
\end{figure}

The complete $H-T$ phase diagram of \Co\ based on our specific heat measurements is displayed in
Fig.~\ref{phase-diagram} for three orientations of the magnetic field, $H \parallel$ [110], $H
\parallel$ [100] (closed and open symbols in panel (a), respectively), and $H \parallel$ [001] (panel (b)).
The second-to-first order change is indicated by $T_0 = 1.1 \pm 0.1$ K for $H
\parallel $ [110], which is about 10\% higher than $T_0$ for $H \parallel$ [100]. The $T_0$ is obtained from the evolution of
the specific heat anomaly and the magneto-caloric data (not shown), with analysis similar to the
one performed for $H \parallel$ [001]~\cite{bianchi:prl-02}. There is anisotropy for the field in
the {\it a-b} plane of \Co~\cite{murphy:prb-02,izawa:prl-01}. This anisotropy is manifested in $=
1.1$\% higher critical field in the [100] direction which develops above $H = 10 $ T, the region of
the first order superconducting transition. The inset (c) of Fig.~\ref{phase-diagram}(a) shows the
evolution of the entropy with magnetic field $H \parallel$ [100] spanning the region of fields from
well into the first order (11.4 T) to well into the second order (8.6 T) regions of the
superconducting phase transition. The entropy is clearly conserved (all curves collapse on a single
curve in the normal state ($T \approx 1.5$ K), proving that in both regimes the specific heat
anomalies are due to the same electrons (and no other degrees of freedom) participating in
superconducting phase transitions. The inset (d) of Fig.~\ref{phase-diagram}(a) shows the magnitude
of the step of the $T_{FFLO}$ anomaly, obtained via the equal entropy construction, as a function
of magnetic field $H \parallel$ [110]. The data are rather linear in field, indicating the tendency
of the anomaly to disappear for fields less then $\approx 9.9$ T. The $T_{FFLO}$ anomaly, indicated
by solid circles for $H \parallel$ [110], also appears to extrapolate towards field close to 10 T
on the $H$-axis. The inset in Fig.~\ref{phase-diagram}(b) shows low temperature electronic specific
heat (Schottky contribution was subtracted) for magnetic field close to $H_{c2} = 4.95$ T with
$H\parallel$ [001]. The low temperature anomaly $T_{FFLO}$ can also be resolved at 4.9 T, 4.875 T,
and 4.85 T. This anomaly was not observed for $ H \le 4.8$ T. $T_{FFLO} = 130$ mK is about half of
the value for $H
\parallel$ [100]. This indicates that FFLO state is more stable when magnetic field is in the $a-b$
plane of this quasi-2D compound, as expected. The tiny high-field/low-temperature corner of the
$H-T$ phase diagram occupied by the FFLO phase for $H\parallel$ [001] is indicated by open
triangles in Fig.~\ref{phase-diagram}(b). The emerging picture therefore is that of a single
$T_{FFLO}$ phase boundary carving out a high field/low temperature part of the superconducting
state of \Co.

A number of theoretical approaches were taken to explore the FFLO state, which resulted in a
variety of possible phase diagrams~\cite{tachiki:zfpb-96,houzet:prb-01,agterberg:jpcm-01}. Our data
is consistent with some of these expectations. The first order superconducting phase transition for
$T_c < T_0$ was predicted by K. Maki~\cite{maki:ptp-64} for Type II superconductor with strong
Pauli limiting. Under these conditions the FFLO state was calculated to occur below the same
temperature $T_0$ for pure superconductors~\cite{gruenberg:prl-66}. Introduction of impurities
modifies this picture: the first order normal-to-superconducting phase transition is expected to be
rather insensitive to the impurity scattering, while the FFLO state is suppressed to lower
temperatures both for the s-wave~\cite{bulaevskii:sjltp-76} and d-wave~\cite{agterberg:jpcm-01}
pairing. \Co\ has been shown to be a d-wave superconductor in a clean
limit~\cite{movshovich:prl-01}, with impurity scattering most likely close to the unitary limit,
based on low temperature thermal conductivity measurements. In such case, a Larkin-Ovchinnikov
state is most likely stabilized in the low temperature/high field corner of the superconducting
state of the $H-T$ phase diagram~\cite{agterberg:jpcm-01}, in accord with our data. Recent Monte
Carlo calculations of the phase diagram of the $d_{x^2-y^2}$ superconductor in magnetic
field~\cite{adachi:cond-mat-03} indicate that the superconducting fluctuations modify the first
order phase transition below $T_0$ into the nearly discontinuous crossover (broadened first order
phase transition), observed experimentally in \Co. These theoretical considerations lead us to
conclude that the $T_{FFLO}$ anomaly is indeed the vortex state - FFLO state phase boundary.

In summary, we have observed the low temperature specific heat anomaly within the superconducting
state of \Co\ in a region of the phase diagram where the normal to superconducting phase transition
is first order, as also demonstrated by the specific heat measurements. This transition is
conclusively identified as due to the formation of the spatially inhomogeneous superconducting FFLO
state, predicted first theoretically about 40 years ago.

We thank L. Boulaevskii, D. Agterberg, K. Maki, Y. Ikeda, and I. Vekhter for stimulating
discussions. Work at Los Alamos National Laboratory was performed under the auspices of the U.S.
Department of Energy. Work at the NHMFL was performed under the auspices of the National Science
Foundation, the State of Florida and the US Department of Energy.


\begin{thebibliography}{26}
\expandafter\ifx\csname natexlab\endcsname\relax\def\natexlab#1{#1}\fi \expandafter\ifx\csname
bibnamefont\endcsname\relax
  \def\bibnamefont#1{#1}\fi
\expandafter\ifx\csname bibfnamefont\endcsname\relax
  \def\bibfnamefont#1{#1}\fi
\expandafter\ifx\csname citenamefont\endcsname\relax
  \def\citenamefont#1{#1}\fi
\expandafter\ifx\csname url\endcsname\relax
  \def\url#1{\texttt{#1}}\fi
\expandafter\ifx\csname urlprefix\endcsname\relax\def\urlprefix{URL }\fi
\providecommand{\bibinfo}[2]{#2} \providecommand{\eprint}[2][]{\url{#2}}

\bibitem[{\citenamefont{Fulde and Ferrell}(1964)}]{fulde-ferrell:pr-64}
\bibinfo{author}{\bibfnamefont{P.}~\bibnamefont{Fulde}} \bibnamefont{and}
  \bibinfo{author}{\bibfnamefont{R.~A.} \bibnamefont{Ferrell}},
  \bibinfo{journal}{Physical Review} \textbf{\bibinfo{volume}{135}},
  \bibinfo{pages}{A550} (\bibinfo{year}{1964}).

\bibitem[{\citenamefont{Larkin and
  Ovchinnikov}(1964)}]{larkin-ovchinnikov:jetp-64}
\bibinfo{author}{\bibfnamefont{A.~I.} \bibnamefont{Larkin}} \bibnamefont{and}
  \bibinfo{author}{\bibfnamefont{Y.~N.} \bibnamefont{Ovchinnikov}},
  \bibinfo{journal}{J. Exptl. Theoret. Phys. (USSR)}
  \textbf{\bibinfo{volume}{47}}, \bibinfo{pages}{1136} (\bibinfo{year}{1964}),
  \bibinfo{note}{[Sov. Phys. JETP {\bf 20}, 762, (1965).]}.

\bibitem[{\citenamefont{Clogston}(1962)}]{clogston:prl-62}
\bibinfo{author}{\bibfnamefont{A.~M.} \bibnamefont{Clogston}},
  \bibinfo{journal}{Phys. Rev. Lett.} \textbf{\bibinfo{volume}{2}},
  \bibinfo{pages}{9} (\bibinfo{year}{1962}).

\bibitem[{\citenamefont{Gruenberg and Gunther}(1966)}]{gruenberg:prl-66}
\bibinfo{author}{\bibfnamefont{L.~W.} \bibnamefont{Gruenberg}}
  \bibnamefont{and} \bibinfo{author}{\bibfnamefont{L.}~\bibnamefont{Gunther}},
  \bibinfo{journal}{Phys. Rev. Lett.} \textbf{\bibinfo{volume}{16}},
  \bibinfo{pages}{996} (\bibinfo{year}{1966}).

\bibitem[{\citenamefont{Berlincourt and Hake}(1963)}]{berlincourt:pr-63}
\bibinfo{author}{\bibfnamefont{T.~G.} \bibnamefont{Berlincourt}}
  \bibnamefont{and} \bibinfo{author}{\bibfnamefont{R.~R.} \bibnamefont{Hake}},
  \bibinfo{journal}{Phys. Rev.} \textbf{\bibinfo{volume}{131}},
  \bibinfo{pages}{140} (\bibinfo{year}{1963}).

\bibitem[{\citenamefont{Kim et~al.}(1965)\citenamefont{Kim, Hempstead, and
  Strnad}}]{kim:pr-65}
\bibinfo{author}{\bibfnamefont{Y.~B.} \bibnamefont{Kim}},
  \bibinfo{author}{\bibfnamefont{C.~F.} \bibnamefont{Hempstead}},
  \bibnamefont{and} \bibinfo{author}{\bibfnamefont{A.~R.}
  \bibnamefont{Strnad}}, \bibinfo{journal}{Phys. Rev.}
  \textbf{\bibinfo{volume}{139}}, \bibinfo{pages}{A1163}
  (\bibinfo{year}{1965}).

\bibitem[{\citenamefont{Shapira and Neuringer}(1965)}]{shapira:pr-65}
\bibinfo{author}{\bibfnamefont{Y.}~\bibnamefont{Shapira}} \bibnamefont{and}
  \bibinfo{author}{\bibfnamefont{L.~J.} \bibnamefont{Neuringer}},
  \bibinfo{journal}{Phys. Rev.} \textbf{\bibinfo{volume}{140}},
  \bibinfo{pages}{A1638} (\bibinfo{year}{1965}).

\bibitem[{\citenamefont{Hake}(1965)}]{hake:prl-65}
\bibinfo{author}{\bibfnamefont{R.~R.} \bibnamefont{Hake}},
  \bibinfo{journal}{Phys. Rev. Lett.} \textbf{\bibinfo{volume}{22}},
  \bibinfo{pages}{865} (\bibinfo{year}{1965}).

\bibitem[{\citenamefont{Maki}(1966)}]{maki:pr-66}
\bibinfo{author}{\bibfnamefont{K.}~\bibnamefont{Maki}}, \bibinfo{journal}{Phys.
  Rev.} \textbf{\bibinfo{volume}{148}}, \bibinfo{pages}{362}
  (\bibinfo{year}{1966}).

\bibitem[{\citenamefont{Gloos et~al.}(1993)\citenamefont{Gloos, Modler,
  Schimanski, Bredl, Geibel, Steglich, Buzdin, Sato, and
  Komatsubara}}]{gloos:prl-93}
\bibinfo{author}{\bibfnamefont{K.}~\bibnamefont{Gloos}},
  \bibinfo{author}{\bibfnamefont{R.}~\bibnamefont{Modler}},
  \bibinfo{author}{\bibfnamefont{H.}~\bibnamefont{Schimanski}},
  \bibinfo{author}{\bibfnamefont{C.}~\bibnamefont{Bredl}},
  \bibinfo{author}{\bibfnamefont{C.}~\bibnamefont{Geibel}},
  \bibinfo{author}{\bibfnamefont{F.}~\bibnamefont{Steglich}},
  \bibinfo{author}{\bibfnamefont{A.}~\bibnamefont{Buzdin}},
  \bibinfo{author}{\bibfnamefont{N.}~\bibnamefont{Sato}}, \bibnamefont{and}
  \bibinfo{author}{\bibfnamefont{T.}~\bibnamefont{Komatsubara}},
  \bibinfo{journal}{Physical Review Letters} \textbf{\bibinfo{volume}{70}},
  \bibinfo{pages}{501 } (\bibinfo{year}{1993}).

\bibitem[{\citenamefont{Huxley et~al.}(1993)\citenamefont{Huxley, Paulson,
  Laborde, Tholence, Sanchez, Junod, and Calemczuk}}]{huxley:jpcm-93}
\bibinfo{author}{\bibfnamefont{A.}~\bibnamefont{Huxley}},
  \bibinfo{author}{\bibfnamefont{C.}~\bibnamefont{Paulson}},
  \bibinfo{author}{\bibfnamefont{O.}~\bibnamefont{Laborde}},
  \bibinfo{author}{\bibfnamefont{J.}~\bibnamefont{Tholence}},
  \bibinfo{author}{\bibfnamefont{D.}~\bibnamefont{Sanchez}},
  \bibinfo{author}{\bibfnamefont{A.}~\bibnamefont{Junod}}, \bibnamefont{and}
  \bibinfo{author}{\bibfnamefont{R.}~\bibnamefont{Calemczuk}},
  \bibinfo{journal}{Journal of Physics: Condensed Matter}
  \textbf{\bibinfo{volume}{5}}, \bibinfo{pages}{7709 } (\bibinfo{year}{1993}).

\bibitem[{\citenamefont{Tenya et~al.}(1999)\citenamefont{Tenya, Yasunami,
  Tayama, Amitsuka, Sakakibara, Hedo, Inada, Haga, Yamamoto, and
  Onuki}}]{tenya:physicab-99}
\bibinfo{author}{\bibfnamefont{K.}~\bibnamefont{Tenya}},
  \bibinfo{author}{\bibfnamefont{S.}~\bibnamefont{Yasunami}},
  \bibinfo{author}{\bibfnamefont{T.}~\bibnamefont{Tayama}},
  \bibinfo{author}{\bibfnamefont{H.}~\bibnamefont{Amitsuka}},
  \bibinfo{author}{\bibfnamefont{T.}~\bibnamefont{Sakakibara}},
  \bibinfo{author}{\bibfnamefont{M.}~\bibnamefont{Hedo}},
  \bibinfo{author}{\bibfnamefont{Y.}~\bibnamefont{Inada}},
  \bibinfo{author}{\bibfnamefont{Y.}~\bibnamefont{Haga}},
  \bibinfo{author}{\bibfnamefont{E.}~\bibnamefont{Yamamoto}}, \bibnamefont{and}
  \bibinfo{author}{\bibfnamefont{Y.}~\bibnamefont{Onuki}},
  \bibinfo{journal}{Physica B} \textbf{\bibinfo{volume}{259-261}},
  \bibinfo{pages}{692 } (\bibinfo{year}{1999}).

\bibitem[{\citenamefont{Norman}(1993)}]{norman:prl-93}
\bibinfo{author}{\bibfnamefont{M.~R.} \bibnamefont{Norman}},
  \bibinfo{journal}{Phys. Rev. Lett.} \textbf{\bibinfo{volume}{71}},
  \bibinfo{pages}{3391} (\bibinfo{year}{1993}).

\bibitem[{\citenamefont{Movshovich et~al.}(2001)\citenamefont{Movshovich,
  Jaime, Thompson, Petrovic, Fisk, Pagliuso, and Sarrao}}]{movshovich:prl-01}
\bibinfo{author}{\bibfnamefont{R.}~\bibnamefont{Movshovich}},
  \bibinfo{author}{\bibfnamefont{M.}~\bibnamefont{Jaime}},
  \bibinfo{author}{\bibfnamefont{J.~D.} \bibnamefont{Thompson}},
  \bibinfo{author}{\bibfnamefont{C.}~\bibnamefont{Petrovic}},
  \bibinfo{author}{\bibfnamefont{Z.}~\bibnamefont{Fisk}},
  \bibinfo{author}{\bibfnamefont{P.~G.} \bibnamefont{Pagliuso}},
  \bibnamefont{and} \bibinfo{author}{\bibfnamefont{J.~L.}
  \bibnamefont{Sarrao}}, \bibinfo{journal}{Phys. Rev. Lett.}
  \textbf{\bibinfo{volume}{86}}, \bibinfo{pages}{5152} (\bibinfo{year}{2001}).

\bibitem[{\citenamefont{Bianchi et~al.}(2002)\citenamefont{Bianchi, Movshovich,
  Oeschler, Gegenwart, Steglich, Thompson, Pagliuso, and
  Sarrao}}]{bianchi:prl-02}
\bibinfo{author}{\bibfnamefont{A.}~\bibnamefont{Bianchi}},
  \bibinfo{author}{\bibfnamefont{R.}~\bibnamefont{Movshovich}},
  \bibinfo{author}{\bibfnamefont{N.}~\bibnamefont{Oeschler}},
  \bibinfo{author}{\bibfnamefont{P.}~\bibnamefont{Gegenwart}},
  \bibinfo{author}{\bibfnamefont{F.}~\bibnamefont{Steglich}},
  \bibinfo{author}{\bibfnamefont{J.~D.} \bibnamefont{Thompson}},
  \bibinfo{author}{\bibfnamefont{P.~G.} \bibnamefont{Pagliuso}},
  \bibnamefont{and} \bibinfo{author}{\bibfnamefont{J.~L.}
  \bibnamefont{Sarrao}}, \bibinfo{journal}{Phys. Rev. Lett.}
  \textbf{\bibinfo{volume}{89}}, \bibinfo{pages}{137002}
  (\bibinfo{year}{2002}).

\bibitem[{\citenamefont{Maki and Tsuneto}(1964)}]{maki:ptp-64}
\bibinfo{author}{\bibfnamefont{K.}~\bibnamefont{Maki}} \bibnamefont{and}
  \bibinfo{author}{\bibfnamefont{T.}~\bibnamefont{Tsuneto}},
  \bibinfo{journal}{Progress of Theretical Physics}
  \textbf{\bibinfo{volume}{31}}, \bibinfo{pages}{945} (\bibinfo{year}{1964}).

\bibitem[{\citenamefont{Tayama et~al.}(2002)\citenamefont{Tayama, Harita,
  Sakakibara, Haga, Shishido, Settai, and Onuki}}]{tayama:prb-02}
\bibinfo{author}{\bibfnamefont{T.}~\bibnamefont{Tayama}},
  \bibinfo{author}{\bibfnamefont{A.}~\bibnamefont{Harita}},
  \bibinfo{author}{\bibfnamefont{T.}~\bibnamefont{Sakakibara}},
  \bibinfo{author}{\bibfnamefont{Y.}~\bibnamefont{Haga}},
  \bibinfo{author}{\bibfnamefont{H.}~\bibnamefont{Shishido}},
  \bibinfo{author}{\bibfnamefont{R.}~\bibnamefont{Settai}}, \bibnamefont{and}
  \bibinfo{author}{\bibfnamefont{Y.}~\bibnamefont{Onuki}},
  \bibinfo{journal}{Phys. Rev. Lett.} \textbf{\bibinfo{volume}{65}},
  \bibinfo{pages}{180504} (\bibinfo{year}{2002}).

\bibitem[{\citenamefont{Murphy et~al.}(2002)\citenamefont{Murphy, Hall, Palm,
  Tozer, Petrovic, Fisk, Goodrich, Pagliuso, Sarrao, and
  Thompson}}]{murphy:prb-02}
\bibinfo{author}{\bibfnamefont{T.~P.} \bibnamefont{Murphy}},
  \bibinfo{author}{\bibfnamefont{D.}~\bibnamefont{Hall}},
  \bibinfo{author}{\bibfnamefont{E.~C.} \bibnamefont{Palm}},
  \bibinfo{author}{\bibfnamefont{S.~W.} \bibnamefont{Tozer}},
  \bibinfo{author}{\bibfnamefont{C.}~\bibnamefont{Petrovic}},
  \bibinfo{author}{\bibfnamefont{Z.}~\bibnamefont{Fisk}},
  \bibinfo{author}{\bibfnamefont{R.~G.} \bibnamefont{Goodrich}},
  \bibinfo{author}{\bibfnamefont{P.}~\bibnamefont{Pagliuso}},
  \bibinfo{author}{\bibfnamefont{J.~L.} \bibnamefont{Sarrao}},
  \bibnamefont{and} \bibinfo{author}{\bibfnamefont{J.~D.}
  \bibnamefont{Thompson}}, \bibinfo{journal}{Phys. Rev. B}
  \textbf{\bibinfo{volume}{65}}, \bibinfo{pages}{100514}
  (\bibinfo{year}{2002}).

\bibitem[{\citenamefont{Shimahara}(1994)}]{shimahara:prb-94}
\bibinfo{author}{\bibfnamefont{H.}~\bibnamefont{Shimahara}},
  \bibinfo{journal}{Phys. Rev. B} \textbf{\bibinfo{volume}{50}},
  \bibinfo{pages}{12760} (\bibinfo{year}{1994}).

\bibitem[{\citenamefont{Hall et~al.}(2001)\citenamefont{Hall, Murphy, Tozer,
  Fisk, Alver, Goodrich, Sarrao, Pagliuso, and Ebihara}}]{hall:prb-01}
\bibinfo{author}{\bibfnamefont{D.}~\bibnamefont{Hall}},
  \bibinfo{author}{\bibfnamefont{E.~C. P. T.~P.} \bibnamefont{Murphy}},
  \bibinfo{author}{\bibfnamefont{S.~W.} \bibnamefont{Tozer}},
  \bibinfo{author}{\bibfnamefont{Z.}~\bibnamefont{Fisk}},
  \bibinfo{author}{\bibfnamefont{U.}~\bibnamefont{Alver}},
  \bibinfo{author}{\bibfnamefont{R.~G.} \bibnamefont{Goodrich}},
  \bibinfo{author}{\bibfnamefont{J.~L.} \bibnamefont{Sarrao}},
  \bibinfo{author}{\bibfnamefont{P.~G.} \bibnamefont{Pagliuso}},
  \bibnamefont{and} \bibinfo{author}{\bibfnamefont{T.}~\bibnamefont{Ebihara}},
  \bibinfo{journal}{Phys. Rev. B} \textbf{\bibinfo{volume}{64}},
  \bibinfo{pages}{212508} (\bibinfo{year}{2001}).

\bibitem[{\citenamefont{Izawa et~al.}(2001)\citenamefont{Izawa, Yamaguchi,
  Matsuda, Shishido, Settai, and Onuki}}]{izawa:prl-01}
\bibinfo{author}{\bibfnamefont{K.}~\bibnamefont{Izawa}},
  \bibinfo{author}{\bibfnamefont{H.}~\bibnamefont{Yamaguchi}},
  \bibinfo{author}{\bibfnamefont{Y.}~\bibnamefont{Matsuda}},
  \bibinfo{author}{\bibfnamefont{H.}~\bibnamefont{Shishido}},
  \bibinfo{author}{\bibfnamefont{R.}~\bibnamefont{Settai}}, \bibnamefont{and}
  \bibinfo{author}{\bibfnamefont{Y.}~\bibnamefont{Onuki}},
  \bibinfo{journal}{Phys. Rev. Lett.} \textbf{\bibinfo{volume}{87}},
  \bibinfo{pages}{057002} (\bibinfo{year}{2001}).

\bibitem[{\citenamefont{Agterberg and Yang}(2001)}]{agterberg:jpcm-01}
\bibinfo{author}{\bibfnamefont{D.~F.} \bibnamefont{Agterberg}}
  \bibnamefont{and} \bibinfo{author}{\bibfnamefont{K.}~\bibnamefont{Yang}},
  \bibinfo{journal}{J. Phys. Condens. Matter} \textbf{\bibinfo{volume}{13}},
  \bibinfo{pages}{9259} (\bibinfo{year}{2001}).

\bibitem[{\citenamefont{Houzet and Buzdin}(2001)}]{houzet:prb-01}
\bibinfo{author}{\bibfnamefont{M.}~\bibnamefont{Houzet}} \bibnamefont{and}
  \bibinfo{author}{\bibfnamefont{A.}~\bibnamefont{Buzdin}},
  \bibinfo{journal}{Phys. Rev. B} \textbf{\bibinfo{volume}{63}},
  \bibinfo{pages}{184521 } (\bibinfo{year}{2001}).

\bibitem[{\citenamefont{Tachiki et~al.}(1996)\citenamefont{Tachiki, Takahashi,
  Gegenwart, Weiden, Lang, Geibel, Steglich, Modler, Paulsen, and
  Onuki}}]{tachiki:zfpb-96}
\bibinfo{author}{\bibfnamefont{M.}~\bibnamefont{Tachiki}},
  \bibinfo{author}{\bibfnamefont{S.}~\bibnamefont{Takahashi}},
  \bibinfo{author}{\bibfnamefont{P.}~\bibnamefont{Gegenwart}},
  \bibinfo{author}{\bibfnamefont{M.}~\bibnamefont{Weiden}},
  \bibinfo{author}{\bibfnamefont{M.}~\bibnamefont{Lang}},
  \bibinfo{author}{\bibfnamefont{C.}~\bibnamefont{Geibel}},
  \bibinfo{author}{\bibfnamefont{F.}~\bibnamefont{Steglich}},
  \bibinfo{author}{\bibfnamefont{R.}~\bibnamefont{Modler}},
  \bibinfo{author}{\bibfnamefont{C.}~\bibnamefont{Paulsen}}, \bibnamefont{and}
  \bibinfo{author}{\bibfnamefont{Y.}~\bibnamefont{Onuki}}, \bibinfo{journal}{Z.
  Phys. B} \textbf{\bibinfo{volume}{100}}, \bibinfo{pages}{369 }
  (\bibinfo{year}{1996}).

\bibitem[{\citenamefont{Bulaevskii and Guseinov}(1976)}]{bulaevskii:sjltp-76}
\bibinfo{author}{\bibfnamefont{L.~N.} \bibnamefont{Bulaevskii}}
  \bibnamefont{and} \bibinfo{author}{\bibfnamefont{A.~A.}
  \bibnamefont{Guseinov}}, \bibinfo{journal}{Sov. J. Low Temp. Phys.}
  \textbf{\bibinfo{volume}{2}}, \bibinfo{pages}{140} (\bibinfo{year}{1976}).

\bibitem[{\citenamefont{Adachi et~al.}()\citenamefont{Adachi, Koikegami, and
  Ikeda}}]{adachi:cond-mat-03}
\bibinfo{author}{\bibfnamefont{H.}~\bibnamefont{Adachi}},
  \bibinfo{author}{\bibfnamefont{S.}~\bibnamefont{Koikegami}},
  \bibnamefont{and} \bibinfo{author}{\bibfnamefont{R.}~\bibnamefont{Ikeda}},
  \bibinfo{note}{cond-mat/0303540}.

\end{thebibliography}

\end{document}